\documentclass[a4paper,12pt]{article}

\usepackage[a4paper]{geometry}
\usepackage[dvipsnames]{xcolor}
\usepackage[utf8]{inputenc} 
\usepackage[hidelinks]{hyperref}
\usepackage{float}
\usepackage{amssymb,amsmath,amsthm}
\usepackage{amsfonts}

\pdfoutput=1
\usepackage{graphicx}
\usepackage{graphicx, float, tabularx, booktabs}
\usepackage{color}
\usepackage{enumerate}
\usepackage[american]{babel}
\usepackage{stackrel}
\usepackage[compress]{cite}
\usepackage[numbers, compress]{natbib}
\usepackage{subcaption}

\usepackage{amsfonts}
\usepackage{epsfig}

\usepackage{graphicx, tabularx}
\newcommand{\beq}{\begin{equation}}
\newcommand{\eeq}{\end{equation}}
\newcommand{\bea}{\begin{eqnarray}}
\newcommand{\eea}{\end{eqnarray}}

\vspace{-0.1cm}
\renewcommand{\thefootnote}{\fnsymbol{footnote}}

\author{Patricio Gaete$^{a}$\thanks{patricio.gaete@usm.cl}, 
Piero Nicolini$^{b,c}$\thanks{nicolini@fias.uni-frankfurt.de} \ and
Euro Spallucci\thanks{spallucci@ts.infn.it}
 \\[1ex]
\small $^a$ Departmento de F\'{\i}sica and Centro
Cient\'{\i}fico-Tecnol\'{o}gico de Valpara\'{\i}so, \\[-0.5ex]
\small  Universidad
T\'ecnica Federico Santa Mar\'{\i}a,  Valpara\'{\i}so, Chile \\[1ex]
\small $^b$ Frankfurt Institute for Advanced Studies (FIAS)\\[-0.5ex]
\small Ruth-Moufang-Str.~1, D-60438 Frankfurt am Main, Germany
\\[1ex]
\small $^c$ Institut f\"{u}r Theoretische Physik, Johann Wolfgang Goethe-Universit\"{a}t Frankfurt\\[-0.5ex]
\small Max-von-Laue-Str.~1, D-60438 Frankfurt am Main, Germany\\
[1ex]
\small $^d$ INFN, Sezione di Trieste, Trieste, Italy
\\[1ex]
}
\date{}
\title{Regularization ambiguity and van der Waals black hole in 2+1 dimensions}

\begin{document}
\maketitle

\vspace{-0.5cm}




\begin{abstract}
\noindent 
{\small 
Charged black holes in a ($2+1$)-dimensional anti-de Sitter space-time suffer from some limitations such as the ambiguity in the definition of the mass and the bad short distance behavior.  In this paper we present a way to resolve such issues. 
By extending the parameter space of the BTZ geometry, we  properly identify the integration constants in order to remove the conical singularity sitting at the origin. In such a way we obtain a well defined Minkowski limit and horizons also in the case of de Sitter background space. On the thermodynamic side, we obtain a proper internal energy, by invoking the consistency with the Area Law, even if the mass parameter does not appear in the metric coefficients. As a further improvement,  we show that it is sufficient to assume a finite size of the electric charge to obtain a short scale regular geometry.  The resulting  solution, generalizing the charged BTZ metric, is dual to a van der Waals gas. 
}
\noindent
    
\end{abstract}




\renewcommand{\thefootnote}{\arabic{footnote}}
\setcounter{footnote}{0}
\thispagestyle{empty}
\clearpage

\section{Introduction}

Despite intense efforts in the last four decades, a full understanding  of the physical origin of the variables associated to black hole thermodynamics is still missing. Semiclassical analysis can, however, still offer a proper platform for understanding different features of the physics of black holes, provided one considers some amendments to the original Hawking-Bekenstein formulation \cite{Bek73,Haw75}. There have been, indeed, two notable developments in the field, namely the role played by noncommutative geometry effects \cite{Wit86,SeW99} and the gauge/gravity duality \cite{Mal99,Wit98b}. 

Noncommutative geometry encodes the intrinsic property of granularity of a quantum spacetime and is connected to the non-local character of string theory \cite{Wit86,SeW99}. Customarily noncommutativity is implemented by using a non-local product, called star product or Moyal product \cite{DoN01}. In the context of black holes, noncommutative effects have mostly been implemented by averaging noncommutative fluctuations on suitable coherent states \cite{SmS03a,SmS03b,SmS04}. Equivalently the sought effects can be obtained by employing another multiplication rule, known as Voros star-product \cite{BGM10}, or by considering a suitable non-local gravity action \cite{MMN11,Nic18}. The attractive feature of noncommutative effects is that they replace the singular behavior of black hole solutions with a regular deSitter region, emerging from the quantum fluctuation of the manifold \cite{NSS06b,Nic09}. Noncommutative geometry also offers intriguing insights in the destiny of a black hole in the last stages of the 
thermal emission \cite{NiW11,NiT11}. Rather than a divergent temperature phase, the black hole undergoes a cooling down phase towards an extremal configuration.

Interestingly, the above scenario  for black holes is common to other paradigms for quantum mechanical black holes, such as the ultraviolet self-completeness \cite{Nic18,NiS12}, the generalized uncertainty principle \cite{IMN13}, string T-duality effects \cite{NSW19}  and other pre-geometric quantum mechanical formulations \cite{SpS15,SpS17}. This would suggest that noncommutative geometry can capture model independent characters.

On the other hand, the gauge/gravity duality is a paradigm emerging  from a line of reasoning started by Bekenstein and Hawking \cite{Bek73,Haw75} and corroborated by the 't Hooft's formulation of the holographic principle \cite{tHo93}. Recent developments from the AdS/CFT correspondence 
have provided valuable insights about the black hole informational content \cite{duel}.  Black hole evolution has to be unitary since they correspond to quantum fields living on the boundary of anti-de Sitter space.
 We further note that  $(2+1)$-dimensional black holes \cite{BTZ92} have also played an important role in such a context. First, they naturally live in an  anti-de Sitter background. Second,  their holographic description allows for the derivation of  the entropy-area law from the  counting of states in a unitary conformal field theory \cite{AGM00}.  

Unfortunately, $(2+1)$-dimensional black holes are not free of problems. It has been noted that the associated thermodynamics, in the presence of a $U(1)$-hair, is ill defined for the arbitrariness of the definition of the black hole mass. The latter can diverge and assume negative values \cite{MTZ00,CaM09}. The issue is connected to the presence of a logarithmic term in the metric coefficient, as a result of the solution of the Maxwell equations in $(2+1)$-dimensions. To this purpose we recall that in previous studies \cite{GaS12,GHS12}, we have considered the effect of   noncommutativity on  static potentials for axionic electrodynamics both in $(3+1)$ and  $(2+1)$ space-time dimensions. Our analyses led to a well-defined noncommutative interaction energy and in both cases we have obtained a fully ultraviolet finite static potential. 
With these ideas in mind, we propose the ultraviolet finiteness of the electrostatic potential as a paradigm to solve the puzzling situation of the $(2+1)$-dimensional black hole thermodynamics.

The paper is organized as follows: in Sec. \ref{sec:btzreview} we review the $(2+1)$-dimensional black hole geometry and we propose a solution to the issue of its thermodynamic variables; in Sec. \ref{sec:uvimproved} we derive a ultraviolet finite $(2+1)$-dimensional black hole solution; in Sec. \ref{sec:concl} we draw the conclusions.

 \section{The standard (charged) BTZ solution: a (critical) appraisal}
\label{sec:btzreview}

The standard Bañados-Teitelboim-Zanelli (BTZ) solution is usually presented as a  vacuum solution of the field equations in $(2+1)$ dimensions \cite{BTZ92}. According to such a derivation, the identification of the integration constants is  a mere formal analogy with the   $(3+1)$ dimensional case.
Some of the shortcomings of the charged BTZ solution have already been  scrutinized in the literature. For instance the mass parameter can diverge and assume negative values; for fixed mass parameter, there is no upper bound on the charge \cite{MTZ00}.
Such inconsistencies of the charged BTZ geometry are associated to the logarithmic profile of the electrostatic potential term. Accordingly, the large distance behavior of the solution can be controlled by introducing a ``large box'' radius \cite{MTZ00}, as a length scale entering the logarithmic term. Such a radius can also be identified with the radius of the AdS background or the radius of the event horizon  \cite{CaM09}. The large box is essentially a regulator that enable the definition of a finite mass and thermodynamic variables.

Conversely,  we are going to propose an alternative solution of the above problems, by solving the  Einstein-Maxwell equations with proper source terms. Thus, a length scale for the electrostatic potential shows up in a transparent way. The proposed method of this section aims to pave the way to the derivation of a ultraviolet finite geometry, whose innovative features are discussed in Section \ref{sec:uvimproved}. 

 We start by the line element
\begin{equation}
 ds^2 = -N^2\left(\, r\,\right) dt^2 + N^{-2}\left(\, r\,\right)dr^2 + r^2 d\phi^2
\end{equation}
describing a $(2+1)$ dimensional space-time, which is solution of the Einstein equations
\begin{equation}
R_{\mu\nu}-\frac{1}{2}\left(\, R-2\Lambda\,\right) g_{\mu\nu}=8\pi G T_{\mu\nu}
\end{equation}
where the gravitational constant in natural units has the
 dimension of a length, \textit{i.e.}, $\left[\,G\,\right]=\mathrm{length}$. The stress tensor
is given in terms of the energy density $\rho$, the radial pressure $p_r$ and the angular
pressure $p_\perp$, namely $ T^\mu{}_\nu=\mathrm{diag}\left(\, -\rho\ ,p_r\ , p_\perp\,\right)$. 
Therefore the Einstein equations read 
\begin{eqnarray}
    && \frac{1}{2r}\frac{dN^2}{dr}=-\Lambda -8\pi G \rho\ ,\label{e1}\\
    &&\frac{1}{2r}\frac{dN^2}{dr}=-\Lambda +8\pi G p_r\ ,\label{e2}\\
    &&\frac{1}{2}\frac{d^2N^2}{dr^2}=-\Lambda +8\pi G p_\perp.\label{e3}
\end{eqnarray}
At this point, we need to carefully discuss the source term. We consider a charged, massive particle, \textit{i.e.}, $M>0$ and $q\neq 0$ sitting at the origin. It is often neglected that, in addition to the charged particle, the BTZ solution contains a topological defect that resembles the Barriola-Vilenkin global monopole \cite{BaV89}, a gravitational object emerging 
also in  nonlocal gravity \cite{KKMN19,Boos20}.  To see this, one can write  the energy density as the sum of the mass term and the electrostatic energy:
\begin{equation}
\rho\equiv \rho_{\rm M} + \rho_{\rm e}\ , \quad \rho_{\rm M}\equiv \frac{M}{2\pi r}\delta(r),
\label{eq:endens}
\end{equation} 
where
 $\rho_{\rm e}$ is given by the $T^{\rm em}_0\ ^0$ component of the energy momentum tensor of the electromagnetic field. 
 Maxwell equations in $(2+1)$ dimensions read
\begin{equation}
 \frac{1}{r}\partial_\mu \left(\, r F^{\mu\nu}\,\right)= J^\nu .
\end{equation}
where, in natural units, $ \left[\, A_\mu\, \right]= L^{-1/2}=\left[\, q\, \right] $. By considering a point-like charge as a source of the electric field
\begin{equation}
J^\nu= \frac{q}{2\pi\, r}\delta\left(\, r\,\right)\, \delta^\nu_0 .
\label{el_current}
\end{equation}
one gets, for a purely electrostatic solutions, $ F^{r0}=E\left(\, r\,\right) $, the following equation 
\begin{equation}
 \frac{1}{r}\partial_r\, \left(\, r\, E\,\right)=\, \frac{q}{2\pi r} \,\delta\left(\, r\,\right).
\end{equation}
Thus, the electric field and the corresponding potential read
\begin{equation}
 E= \frac{q}{2\pi r}\ ,\quad \phi = -\frac{q}{2\pi}\,\ln\left(\, \frac{r}{r_0}\,\right)
\end{equation}
where, $r_0$ is an arbitrary integration constant $r_0$.  We notice that the electric potential  is logarithmic divergent both at large and small length scales, a feature that has important consequences in the presence of gravity. 
The electrostatic energy density  is then given by
\begin{equation}
\rho_{\rm e}=-T^{\rm em}_0\ ^0= \frac{1}{2}\, E^2= \frac{q^2}{8{\pi}^2 r^2}\ .
\end{equation}
When trying to integrate the Einstein equations, 
\begin{equation}
N^2(r)= -2\Lambda \int^r  dr^\prime r^\prime -16\pi G\int^r  dr^\prime r^\prime \rho,
\label{eq:prellineel}
\end{equation}
one encounters a problem: The second term gives again a logarithmic divergent contribution both for $r\to 0$ and $r\to \infty$. Thus, one must cut-off the integration
range at some length scale $r_0$. The resulting metric is
\begin{equation}
 N^2(r)=N^2(r_0)+ \frac{r^2}{l^2} -\frac{r_0^2}{l^2} - \frac{G\,q^2}{\pi}\ln\frac{r^2}{r_0^2} .
\label{chbtz}
\end{equation}
where $\Lambda\equiv -1/l^2$.\\
The above solution depends on two constants, namely, $r_0$ and $N^2(r_0)$. To obtain the BTZ geometry one has to introduce a mass term\footnote{Note that, despite having the same symbol, the mass term in \eqref{eq:btzparam} has a different origin from that in \eqref{eq:endens}. In \eqref{eq:btzparam} the mass corresponds to an integration constant.} by assuming
\begin{equation}
8GM  \equiv \frac{r_0^2}{\ell^2} -N^2(r_0). 
\label{eq:btzparam}
 \end{equation}
The above relation can also be thought as a setting of the parameter $r_0$ in terms of an implicit  function of $M$ and $N^2(r_0)$, namely $r_0=r_0\left(M, N^2(r_0)\right)$. From this viewpoint, gravity provides a sort of dynamical regulator  of the logarithmic divergence of the electrostatic potential. As a result one can display the standard BTZ geometry:
\begin{equation}
 N^2= \frac{r^2}{l^2} -8GM - \frac{G\,q^2}{\pi}\ln\frac{r^2}{r_0^2}.
\label{chbtz2}
\end{equation}
On the geometrical side, the mass term corresponds to a conical singularity in $r=0$ for $q=0$. In the charged case, the electrostatic term worsens the situation by introducing an additional genuine curvature singularity  at the origin.  Another feature, which has largely been  ignored in the literature, apart from few exceptions (see \textit{e.g.} \cite{MTZ00,CaM09}), is that the parameter $M$ can  always be re-scaled out due to the arbitrariness of the integration constant $r_0$, by setting
 \begin{equation}
 r_0^2\longrightarrow r_0^2 \exp\left(\, \frac{8 \pi M}{ q^2}  \,\right).\ \label{rescaling}
\end{equation}
From a physical viewpoint this is the consequence of the degree of divergence of the electromagnetic energy in $(2+1)$-dimensions. The above change of the integration constant is a gauge freedom of the electromagnetic field and does not change its contribution to the matter action. 
  Eq. \eqref{rescaling}, however, opens the question of  the proper interpretation  of the solution, \textit{e.g.}, the existence of horizons and their thermodynamics, in case the metric does not explicitly present a mass term among its parameters.

\begin{figure}[ht]
\begin{center}
\includegraphics[width=0.5\textwidth]{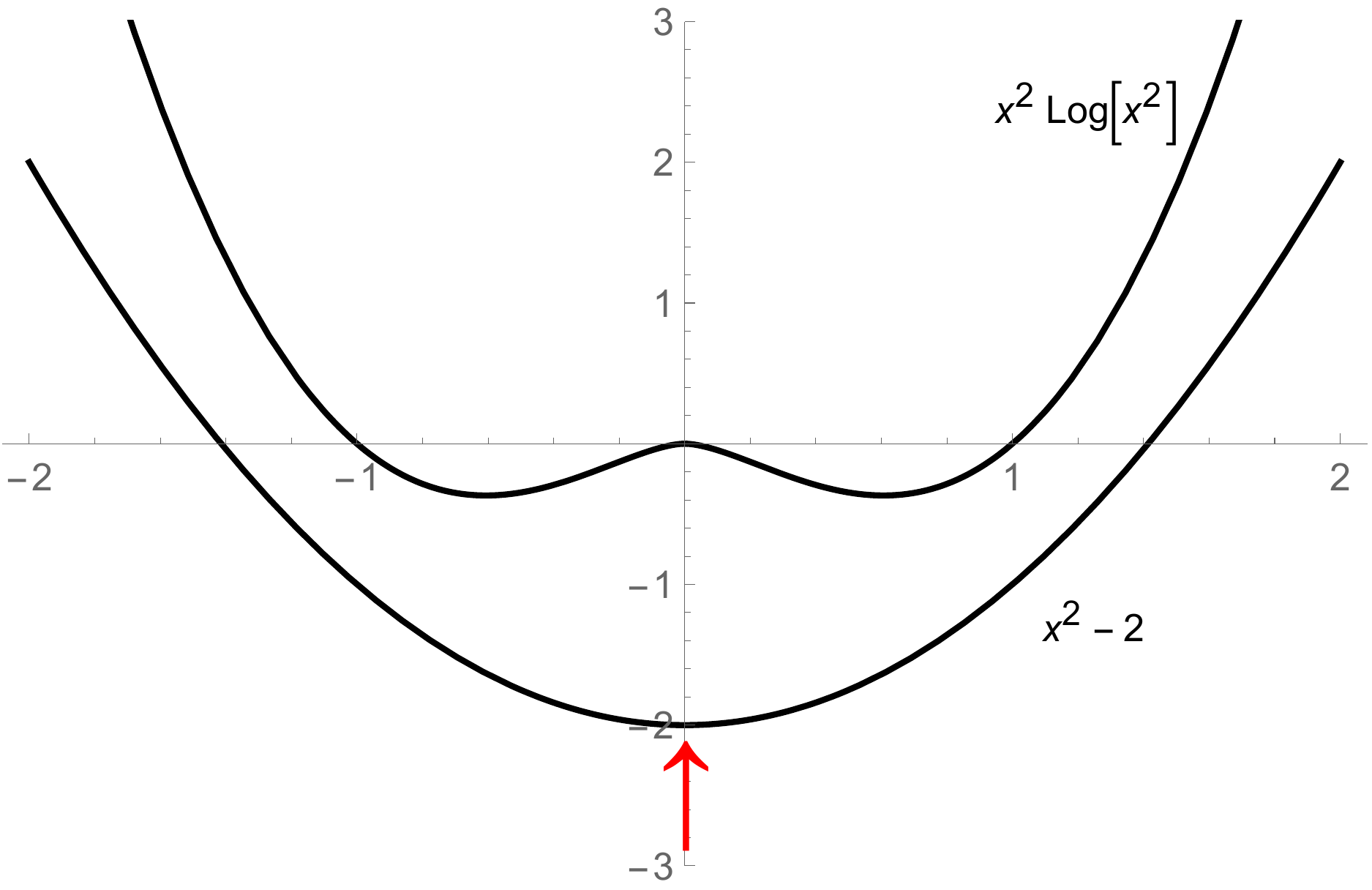} 
\end{center}
\caption{\small Zeros of the function $N^2\left(\, x\,\right)$ in \eqref{btz1} for $N^2_0=-2$ can be obtained from the plot of the functions $x^2-2$ and $x^2\ln x^2$. In the above label $\log$ stands for $\ln$. By increasing $N^2_0$ the parabola $x^2+N^2_0$ is lifted along the direction of the arrow and one or two intersections on the positive real axis can develop.}
\label{fig:horizonequationBTZ}
\end{figure}

We notice that, irrespective of the value of $r_0$ and  $N^2(r_0)$, the function $N^2(r)$ in \eqref{chbtz} is the sum of a monotonically increasing and a monotonically decreasing function, resulting in a convex function with a \emph{single}, local, minimum given by  
\begin{equation}
\frac{dN^2}{dr}= 0 \longrightarrow r_\mathrm{min}^2 = \frac{G q^2 l^2}{\pi}.\label{rex}
\end{equation} 
Every choice of $r_0$ and $N^2(r_0)$ actually corresponds to setting the height of such a minimum. One can see that, in such a local minimum, the metric function takes the value
\begin{equation}
 N^2\left(\, r_\mathrm{min}\,\right)= N^2(r_0)-\frac{r_0^2}{l^2} + \frac{r_\mathrm{min}^2}{l^2}\left(1-\ln\frac{r_\mathrm{min}^2}{r_0^2}\right).
\end{equation} 
The minimum can be negative, vanishing or positive,  corresponding to the presence of two horizons, $r_\pm$,  a Cauchy horizon, $r_-$, and an event horizon, $r_+$; one
(degenerate) horizon or no horizons. The degenerate case corresponds to an extremal black hole, occurring for the coalescence of the two horizons, namely $r_\mathrm{extr}\equiv  r_-=r_+=r_\mathrm{min}$. 
This can be seen by writing $N^2\left(\, r_\mathrm{min}\,\right)$ as
\begin{equation}
 N^2\left(\, x\,\right)=\frac{r_0^2}{l^2}\left( N^2_0+ x^2-x^2\ln x^2 \right),
\label{btz1}
\end{equation} 
with $x\equiv r_\mathrm{min}/r_0$ and $N^2_0\equiv (l^2/r_0^2) N^2(r_0)-1$ (see Fig. \ref{fig:horizonequationBTZ}).

For $N^2_0<-1$, the minimum is always negative, $N^2\left(\, x\,\right)<0$, corresponding to two horizons, $r_\pm $, irrespective of the value of $q$.

For $N^2_0=-1$, the minimum vanishes only for $r_\mathrm{min}=r_\mathrm{extr}=r_0$, namely for $|q|=(r_0/l)\sqrt{\pi/G}$, corresponding to an extremal configuration. For  $|q|\neq(r_0/l)\sqrt{\pi/G}$ the minimum is negative, corresponding to two horizons, $r_\pm$.

For $-1<N^2_0<0$, the minimum  vanishes for two values of the minimum radius $r_\mathrm{min}$, leading to extremal configurations at $r_\mathrm{extr}=r_1<r_0$ and  $r_\mathrm{extr}=r_2>r_0$, corresponding to two distinct values of the charge parameter, $q_1$ and $q_2$. The minimum is negative for $q<q_1$ or for $q>q_2$, corresponding to two horizons, $r_\pm$. For $q_1<q<q_2$, the minimum is positive and no horizons form.  In Fig. \ref{fig:horizonequationBTZ} the absence of horizons corresponds to the portion of the curve $x^2\ln x^2$ lying below the parabola, when the latter is lifted in the direction of the arrow.

For $N^2_0=0$, one finds a borderline situation of the previous case, with  $q_1\to 0$ corresponding to $r_1=0$, and $q_2\to q^\ast$. As a result, there exists only a nonvanishing value of $q^\ast$ such that the extremal configuration is realized. For larger charge, $q>q^\ast$, the minimum is negative corresponding to two horizons, $r_\pm$. For $q<q^\ast$ horizons do not form.

Finally for $N^2_0>0$, there is  only a nonvanishing value of $q$ such that the extremal configuration is realized. For larger $q$, the minimum is negative corresponding to two horizons, $r_\pm$. For smaller charge, horizons do not form.

We note that, by studying the metric for a generic $N_0^2$, we have extended the parameter space governing the horizon structure. Since $N_0^2$ has no actual physical meaning, it can assume any value. On the other hand the parametrization proposed by BTZ in \eqref{eq:btzparam} limits the analysis to the case $N_0^2<0$, being the mass parameter positive defined. To this purpose we recall that an extension of the parameter space have already been obtained in \cite{CaM09} within the proposal of interpreting the BTZ black hole entropy in terms of a Cardy formula for the two dimensional dual conformal field theory.

In the case the metric admits an event horizon, $r_+$, we can define a temperature. 
According to Hawking this is given by 
\begin{equation}
T_{\rm H} 
= \frac{r_+}{2\pi l^2}\left(\, 1 -\frac{r_\mathrm{extr}^2}{r_+^2}
\,\right)\label{th}.
\end{equation}
Not surprisingly, the temperature vanishes at the extremal configuration, $r_+=r_\mathrm{extr}$.

Conventionally the mass term coincides with the internal energy, namely the quantity $U=M(r_+, q)$ one obtains by solving the horizon equation, $N^2(r)=0$. Then, one can insert (\ref{th}) in the First Law, $dU= T_{\rm H} dS+\phi dq$, to obtain the entropy $S$. This is, however, the case only if a mass term is explicitly present in the metric coefficient, thing that does not, in general, occur for  a $(2+1)$-dimensional charged black hole geometry. Due to the freedom of the parameter $r_0$, there is no mass parameter that can be the candidate for the role of internal energy $U$.
The situation might appear similar to the case of the Rindler geometry that represents the spacetime in the presence of an inertial field.  For  Rindler geometries, however, it is the acceleration that determines the existence of the horizon and there is no actual mass term irrespective of the parametrization.

 As a result one has to figure out an alternative way to define the internal energy without relying on the parametrization in \eqref{eq:btzparam}. As a solution of the issue, one can invoke the Area Law, namely the fact that the entropy of gravitational systems have a holographic nature. 
Accordingly,  we propose the following form for $U$
\begin{equation}
 U\equiv \frac{1}{8G}\left(\,
\frac{r_+^2}{l^2}-\frac{r_\mathrm{extr}^2}{l^2}  -\frac{r_\mathrm{extr}^2}{l^2}\, \ln\frac{r_+^2}{r_\mathrm{extr}^2}\,\right)
\label{uguess}
\end{equation}
that is compatible with the Area law, \textit{i.e.}, 
\begin{equation}
dS= \frac{\pi}{2G} dr_+ \rightarrow S =\frac{\pi}{2G}\int{r_\mathrm{extr}}^{r_+} dr_+' = \frac{1}{4G}\left(\, A_+ -A_\mathrm{extr}\,\right),
\label{alaw}
\end{equation}
where $A_+\equiv 2\pi r_+$ and $A_\mathrm{extr}\equiv 2\pi r_\mathrm{extr}$. In other words, it is possible to consistently define the internal energy of the event horizon for any $N_0^2$, despite no mass term enters the metric coefficient.
This feat is in agreement with what found in \cite{CaM09}, even if with important differences. First,  the proposed method of derivation is original and alternative to the usual  regularization of the mass with a large box along the lines of \cite{MTZ00}. Second, the above thermodynamic variables vanish at the extremal configuration, namely for $T=0$ one has $U=S=0$ corresponding to a frozen system without statistical ignorance. Such an important property is not captured in \cite{CaM09}, whose entropy never vanishes due to a specific choice of integration constants. Residual entropies at zero temperature can exist as the effect of state degeneration in some condensed matter systems such as carbon monoxide and spin ice. A similar interpretation for black holes is not known and there is no comment in \cite{CaM09} about such a potential issue.

In order to determine if the above metric describes a system that  is ``dual'' to some kind of fluid, we consider the cosmological constant as a dynamical variable playing the role of a pressure \cite{Dol11,KuM12,FMM15,KuM17}.
Along this line of reasoning we write the AdS vacuum
equation of state as 
\begin{equation}
P=-\rho_{\mathrm{AdS}}=\frac{1}{8\pi G l^2}\ \label{pressure}
\end{equation}
where $\rho_{\mathrm{AdS}}\equiv\Lambda/(8\pi G)$. 
This equation says that the vacuum compresses the black hole by exerting a positive pressure.

We recall that in $(2+1)$-dimensions, the ``volume'' of the black hole is simply given by the area of the ``black disk'':
\begin{equation}
 V=\pi r_+^2 .\label{volume}
\end{equation}
We can, therefore,  write the equation of state for the fluid which is dual to the black hole. By expressing $1/l^2$  in $T_{\rm H}$  in terms of $P$ defined 
in equation (\ref{pressure}), we find
\begin{equation}
 T_{\rm H}= 4G r_+ P -\frac{1}{2\pi^2}\frac{G q^2}{r_+}.
\end{equation}
 Now, if we define the \emph{specific volume } of the fluid as
 \begin{equation}
  v\equiv \frac{V}{{\cal N}} = 4G r_+ , \label{vspec}
 \end{equation}
where ${\cal N}$ accounts for the fluid degrees of freedom, 
we obtain the following equation of state 
\begin{equation}
 P= \frac{T_{\rm H}}{v} +\frac{2G^2 q^2}{\pi^2v^2}\label{gperf}.
\end{equation}
Eq. (\ref{gperf}) describes a \emph{perfect gas} with a short distance correction due to the electrostatic 
repulsion.
A first interesting result is that the specific volume cannot be arbitrary small since $r_+\ge r_\mathrm{extr} $. Accordingly we can define minimum value
of $v$ as $v_{\mathrm{min}}\equiv 4G r_\mathrm{extr}$.
We notice that, for any $T_{\rm H}$, the  function $P=P(v)$ is monotonically decreasing and admits a global maximum value
\begin{equation}
P_{\mathrm{max}}= P(v_{\mathrm{min}})= \frac{T_{\rm H}}{4G r_\mathrm{extr}} +\frac{q^2}{8\pi^2r_\mathrm{extr}^2}
=-\rho_{\mathrm{AdS}}\left(\, 1 +\frac{2\pi^{3/2}lT_{\rm H}}{q\sqrt{G}}\,\right)
\end{equation}
The conclusion is that there are no phase transitions.

We stress here that the above analysis in AdS vacuum coincides with the results presented in \cite{GKM12}.  This is an additional piece of evidence that the proposed method leads to consistent results.

\section{Ultraviolet improved solution}
\label{sec:uvimproved}

From the previous section it has emerged that the solution \eqref{chbtz} presents some pathology. The presence of a gravitational monopole at the origin introduces a conical singularity. 
We also learned that the mass term is an irrelevant parameter in the solution. 
We recall here that noncommutative geometry has already been considered in lower dimensional black hole geometries \cite{MuN11,MyY09,Par09,RKB13,Lia12,TeL12}.
In all such cases  the noncommutative smearing has been applied to the mass term much in the same way of what done in the higher dimensional cases \cite{NSS06b,Nic09,Riz06,SSN09,SmS10,MoN10b,NiS10}. Against this background,  the results of the previous section show that it is possible to obtain a well defined thermodynamics without  the mass term. In the present section we aim to show that the smearing of the mass term is not even necessary to obtain a short scale regular solution. The only genuine short scale singularity is due to the electrostatic term. To this purpose we recall that the derivation of a regular charged  BTZ metric has already been addressed in the literature by introducing a topological Chern-Simon term to the gauge field action \cite{Cle95}. The net result has been an horizonless geometry. Our proposal, however, departs from such a result. In what follows we aim improve the charged BTZ geometry without preventing it from admitting horizons and without any modification of the gauge field action. 

We start by addressing the issue of the conical singularity, that is present also in the absence of the charge, $q=0$. Up to now we considered  $N^2_0$ as a free parameter emerging from the integration of Einstein equations. It is, however, possible to set the value of $N^2_0$ by following a procedure analogue to higher dimensional black hole solutions. In the absence of mass and charge, the spacetime has to match a regular geometry with constant curvature. This means that  in $(2+1)$-dimensions, one can and has to eliminate the deficit angle in order to have a physically consistent solution. In other words one has to require that
\begin{equation}
N^2(r_0)-\frac{r_0^2}{l^2}= 1.
\label{eq:Nzerosetup}
\end{equation}
After this is done, one can set $1/l^2=0$ to find the Minkowski limit. 
We note that the removal of the monopole term by means of \eqref{eq:Nzerosetup} is possible only because we extended the parameter space of the solution. Within the BTZ parametrization \eqref{eq:btzparam} there is no way to fulfill the condition \eqref{eq:Nzerosetup} for $M=0$. In other words the Minkoswki limit is never attainable within the BTZ proposal.

As a result the metric \eqref{chbtz} can be written as
\begin{equation}
N^2(r)=1 +  \frac{r^2}{l^2}  - \frac{G\,q^2}{\pi}\ln\frac{r^2}{r_0^2} .
\end{equation}
We note that monopole has been removed but $r_0$ can still change within the set of values determined by the condition \eqref{eq:Nzerosetup}. It is therefore convenient to introduce  again  a dimensionless variable $u=r/r_0$ to get:
\begin{equation}
N^2(u)=1 -\lambda u^2  - \tilde{q}^2\ln u^2,
\end{equation}
where $\lambda \equiv - r_0^2 / l^2$ and $\tilde{q}^2\equiv G\,q^2/\pi$ are dimensionless constants.

 At this point we note that for $\tilde{q}=0$, there are horizons only if the cosmological term is positive. In other words, the necessity of having an anti-de Sitter background to fulfill the horizon equation holds only for the BTZ parametrization \eqref{eq:btzparam}. For the rest of the paper we keep, however, the standard anti-de Sitter background, $\lambda <0$, for the ease of the presentation.

As a second step of the discussion, one needs to address the issue of the electrostatic term.
We recall here that a point like charge is just an ideal mathematical model. From a physical viewpoint, one actually expects the charge to be distributed with a certain profile of width $a$.   The latter parameter can be  thought
as a ultraviolet cut off of a given theory or simply as the characteristic length scale of the system under consideration. For instance,  in the case of a proton the width is of the order of the nuclear radius,  $a\sim 1$ fermi. For the present problem, we have already an array of length scales, such as $r_0$, $l$, $G$ and $q^{-2}$, but we assume $a$ as an additional independent parameter to keep the discussion general. 

At large distances, $r\gg a $,  the charge distribution appears like a peaked distribution. At scales of the order of $a$, however, the profile of the charge distribution is visible.
Since the main purpose  of this section is to show how a finite width charge distribution can remove the curvature singularity in $r=0$, we consider the case of a Gaussian distribution, even if other distributions may be equally motivated and physically consistent. As a result we start from
\begin{equation}
J^0(r)  =   \frac{q}{4\pi a^2 }e^{ -r^2/4a^2 }. \label{BH40}
\end{equation}
By solving the Maxwell equations, the radial component of the electric field  $E(r)$ reads 
\begin{equation}
E(r) = \frac{q}{2\pi r}\gamma\left(\, 1\ ; r^2/4a^2\,\right)\ . \label{BH45}
\end{equation}
Here the lower incomplete Gamma function is defined as
\begin{equation}
\gamma \left(\, \alpha\ ; x\, \right) = \int_0^x {dt} \ t^{\alpha - 1} e^{ - t}\ , \label{BH35}
\end{equation}
%
that can be written, for $\alpha=1$, in terms of elementary
functions as:
\begin{equation}
\gamma \left(\, 1\ ; x\, \right) =  1- e^{-x}.
\label{eq:gamma1}
\end{equation}
Eq. \eqref{BH45} and \eqref{eq:gamma1}  say that, for $r\to 0$,  the electric field is linearly vanishing and 
the electrostatic energy density, $\rho_{\rm e}$, is  quadratically  vanishing. By inserting the total energy density in \eqref{e1} and integrating the field equations as before , we find
\begin{equation}
 N^2(r)=N^2(r_0)+ \frac{r^2}{l^2} -\frac{r_0^2}{l^2} - \frac{2G\,q^2}{\pi}F\left(\, r\ , r_0\ , a\,\right),
 \label{eq:regmetric1}
\end{equation}
where  the function $F$ is defined as
\begin{equation}
 F\left(\, r\ , r_0\ , a\,\right)\equiv \int_{r_0}^{r} \frac{d{t}}{{t}} \left(\, 1 -2 e^{-{t}^2/4a^2} +e^{-{t}^2/2a^2} \right).
 \label{eq:integralF}
\end{equation} 
 The upper bound of the above integral provides the dependence on the radial coordinate $r$. The lower bound  provides only a  constant depending on $r_0$. Up to such a constant, one finds that the asymptotic behaviors of the function $F$ are:
\begin{equation}
F\left(\, r\ , r_0\ , a\,\right)\sim \ln(r/r_0)\ \mathrm{for}\ r\gg a
\end{equation}
and 
\begin{equation}
F\left(\, r\ , r_0\ , a\,\right)\sim \frac{r^4}{64a^4} \ \mathrm{for}\ r\ll a.
\end{equation}
As expected, the curvature singularity in $r = 0$ has been improved
 in favor of a harmless conical one, that can be removed by a suitable choice of the integration constants. The logarithmic divergence appears only at large distance $r \gg a$, where the actual
width of the charge distribution cannot be resolved.

The integral in \eqref{eq:integralF} can be solved analytically and reads
\begin{eqnarray}
F\left(\, r\ , r_0\ , a\,\right)&=& \ln(r/r_0)-\frac{1}{2}\ \Gamma\left(\, 0\ ; r^2/2a^2\, \right)\\+\ \Gamma\left(\, 0\ ; r^2/4a^2\, \right) &-&\Gamma\left(\, 0\ ; r_0^2/4a^2\, \right)+\frac{1}{2}\ \Gamma\left(\, 0\ ; r_0^2/2a^2\, \right), \nonumber
\end{eqnarray}
where 
\begin{equation}
\Gamma\left(\, 0\ ; x\, \right)=\int_x^\infty \frac{dt}{t}\ e^{ - t}
\end{equation}
is the upper incomplete gamma function. One can calculate the value of  $F$ at $r=0$ and find a finite constant $F_0$, namely
\begin{equation}
F_0=-\frac{\gamma}{2}+\ln\left(\frac{4a}{r_0\sqrt{2}}\right)
-\Gamma\left(\, 0\ ; r_0^2/4a^2\, \right)+\frac{1}{2}\ \Gamma\left(\, 0\ ; r_0^2/2a^2\, \right)
\end{equation}
with $\gamma$ the Euler-Mascheroni constant. The function $F\left(\, r\ , r_0\ , a\,\right)$ describes a short scale regularized electrostatic potential. 
The presence of an effective size $a$ resembles the charge screening effect observed in the context of $(2+1)$-noncommutative electrodynamics \cite{GHS12}. 

We have already noticed that the electrostatic energy vanishes at $r=0$. The cosmological term dies off too, being proportional to $r^2$. As a result in the vicinity of the origin there is no  mass-energy content able to curve the spacetime that results locally flat. Accordingly, one can require that  $N^2(0)=1$ to get rid of the monopole term. This implies the following new condition for the integration constants and the parameters of the solution: 
\begin{equation}
 N^2(r_0) -\frac{r_0^2}{l^2}= 1 + \frac{2G\,q^2}{\pi}F_0\left(\, r_0\ , a\,\right).
 \label{eq:Nzerosetupcharged}
\end{equation}
The above equation is the  extension of \eqref{eq:Nzerosetup} to the charged case.
At this point one can write the short scale regular charged black hole metric as
\begin{equation}
 N^2(r)=1+ \frac{r^2}{l^2} - \frac{2G\,q^2}{\pi}\Big( F\left(\, r\ , r_0\ , a\,\right)-F_0\Big). 
 \label{eq:regmetric2}
 \end{equation}
Once \eqref{eq:Nzerosetupcharged} is fulfilled the form of the above metric coefficient cannot be modified by changing the value of $r_0$. Again it is useful to introduce a dimensionless variable $u=r/r_0$ and display the metric as
\begin{equation}
 N^2(u)=1-\lambda u^2 - 2\tilde{q}^2\Big(F\left(\, u\ , \tilde{a}\,\right)-F_0(\tilde{a})\Big).
 \label{eq:regmetricdimensionless}
\end{equation}
where $\tilde{a}=a/r_0$. 

\begin{figure}[ht]
\begin{center}
\includegraphics[width=0.5\textwidth]{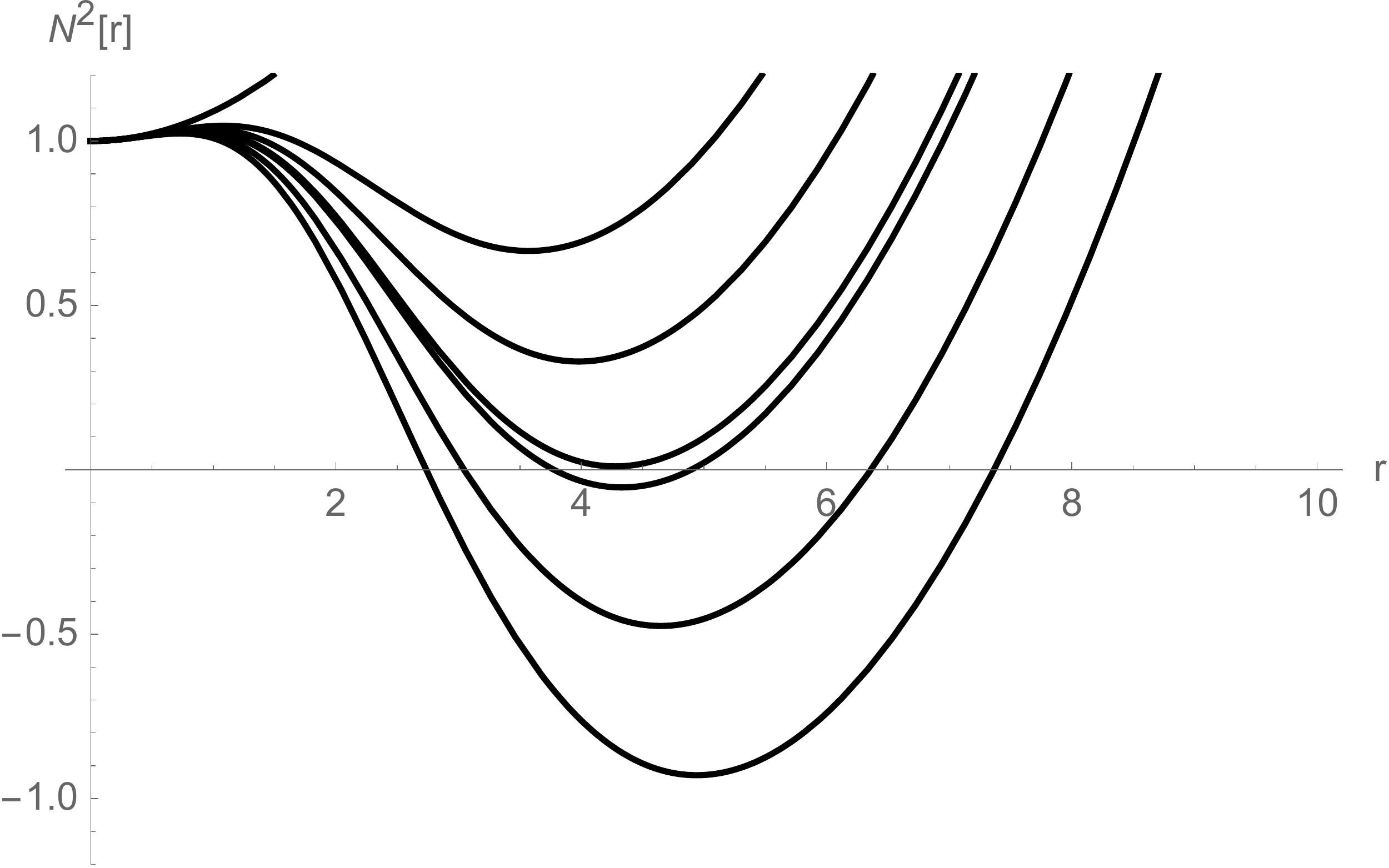} 
\end{center}
\caption{\small The metric function $N^2(r)$ from \eqref{eq:regmetricdimensionless}. Curves are displayed by increasing the charge parameter from the upper curve to the lower curve. We assumed $\tilde{q}=0$, $1.20$, $1.45$, $1.66$, $1.70$, $1.95$ and $2.20$, $\lambda=-0.09$ and $\tilde{a}=0.91$. The case $\tilde{q}=0$ corresponds to the parabola in the left upper part.\label{FIGa}}
\end{figure}

To study the horizon equation we start by noticing that leading correction to Minkowski space in a neighborhood of the origin is dominated by  the monotonically increasing cosmological term, $N(r\approx 0)\sim 1+ r^2/l^2$ being the electrostatic term quartically dependent on the radial coordinate  $\sim r^4/a^4$ as far as $r\ll a$. At large scales the cosmological term is again dominant since it diverges quadratically versus the logarithmic divergence of the electrostatic term $-q^2\ln(r/r_0)$. As a result one has $N^2(r\to\infty)\sim r^2/l^2$.  Between the two regimes there is room for negative contributions coming from the electrostatic term. 
As a result the function $N^2(r)$ has a local maximum at short scale, and a minimum, at intermediate scales. 
 One can easily see this by calculating the second derivative of the metric coefficient at short and large scales:
\begin{eqnarray}
\frac{d^2(N^2)}{dr^2}&\approx & \frac{2}{l^2}-\frac{3}{8}\frac{Gq^2}{\pi}\frac{r^2}{a^4}\ \mathrm{for} \ r\ll a \label{eq:seconddNshort}\\
&\approx & \frac{2}{l^2}+\frac{2Gq^2}{\pi r^2} \ \mathrm{for} \ r\gg a \label{eq:seconddNlong}
\end{eqnarray}
Eq. \eqref{eq:seconddNshort} indicates that the unique stationary point at short scales coming from the equation $dN^2/dr=0$ is a local maximum, being $d^2(N^2)/dr^2=-4/l^2<0$ there. The function $N^2(r)$ can have a root for $r\ll a$ only for large charge, namely only if $q^2>64 (\pi/Gl^2)a^2$. This means that a Cauchy horizon forms. Otherwise for smaller charge, the function $N^2$ crosses the $r$-axis from above at $r>a$ or does not cross the $r$-axis at all.    Eq. \eqref{eq:seconddNlong} says that for $r\gg a$ the function $N^2(r)$ admits a minimum.  When the additional condition $N^2(r_\mathrm{min})=0$ is fulfilled,  such a minimum occurs at the  black hole extremal radius,  $r_\mathrm{extr}=r_\mathrm{min}>a$.
The horizon structure is therefore similar to what seen in Sec. \ref{sec:btzreview}, namely, two horizons $r_\pm$, one degenerate horizon $r_\mathrm{extr}$ or no horizon -- see Fig. \ref{FIGa}. The latter case corresponds to a regular horizonless geometry and not to a naked singularity.  

On the thermodynamic side, the black hole temperature reads: 
\begin{equation}
T_{\rm H}=\frac{r_+}{2\pi l^2}\left[1-\frac{r_\mathrm{extr}}{r_+}\frac{F'(r_+)}{F'(r_\mathrm{extr})}\right]
\label{eq:reg_temp}
\end{equation}
Given the relation 
\begin{equation}
\frac{Gq^2 l^2}{\pi}=\frac{r_\mathrm{extr}}{F^\prime (r_\mathrm{extr})},
\end{equation}
one finds
\begin{equation}
T_{\rm H}=\frac{r_+}{2\pi l^2}
\left[\, 1 -  \frac{Gq^2 l^2}{\pi} \frac{1}{r^2_+} \left(\, 1 -2 e^{-r_+^2/4a^2} +e^{-r_+^2/2a^2}\,\right)   \,\right].
\label{eq:reg_temp2}
\end{equation}
Being $a<r_\mathrm{extr}\leq r_+$, the corrections due to size of the source $a$ are small. The leading term of the temperature coincides with what found in \eqref{th}.

As from the discussion in Sec. \ref{sec:btzreview}, there is the problem of defining the internal energy of the system in the absence of an explicit mass term. To solve the puzzle, we again invoke the compatibility with Area law, $dU= T_{\rm H} dS+\phi dq$,  to find
\begin{equation}
U=\frac{1}{8G}\left\{\frac{r_+^2}{l^2}-\frac{r_\mathrm{extr}^2}{l^2}-\frac{2r_\mathrm{extr}}{l^2F'(r_\mathrm{extr})}\Big[F(r_+)-F(r_\mathrm{extr})\Big]\right\}.
\label{eq:reg_int_en}
\end{equation}
where the electric potential is now $\phi=-(q/2\pi) \Big[F(r)-F_0\Big]$.

We note that temperature in \eqref{eq:reg_temp} and \eqref{eq:reg_temp2} correctly reproduces the classical result, \textit{i.e.} \eqref{th}, in the large distance limit, $r\gg a$, while it vanishes as $r_+\to r_\mathrm{extr}$. 
Such two asymptotic behaviors are also common to the internal energy \eqref{eq:reg_int_en}, \textit{i.e.}, it approaches the classical result at large distance and vanishes at the extremal configuration.

\begin{figure}[ht]
\begin{center}
\includegraphics[width=0.5\textwidth]{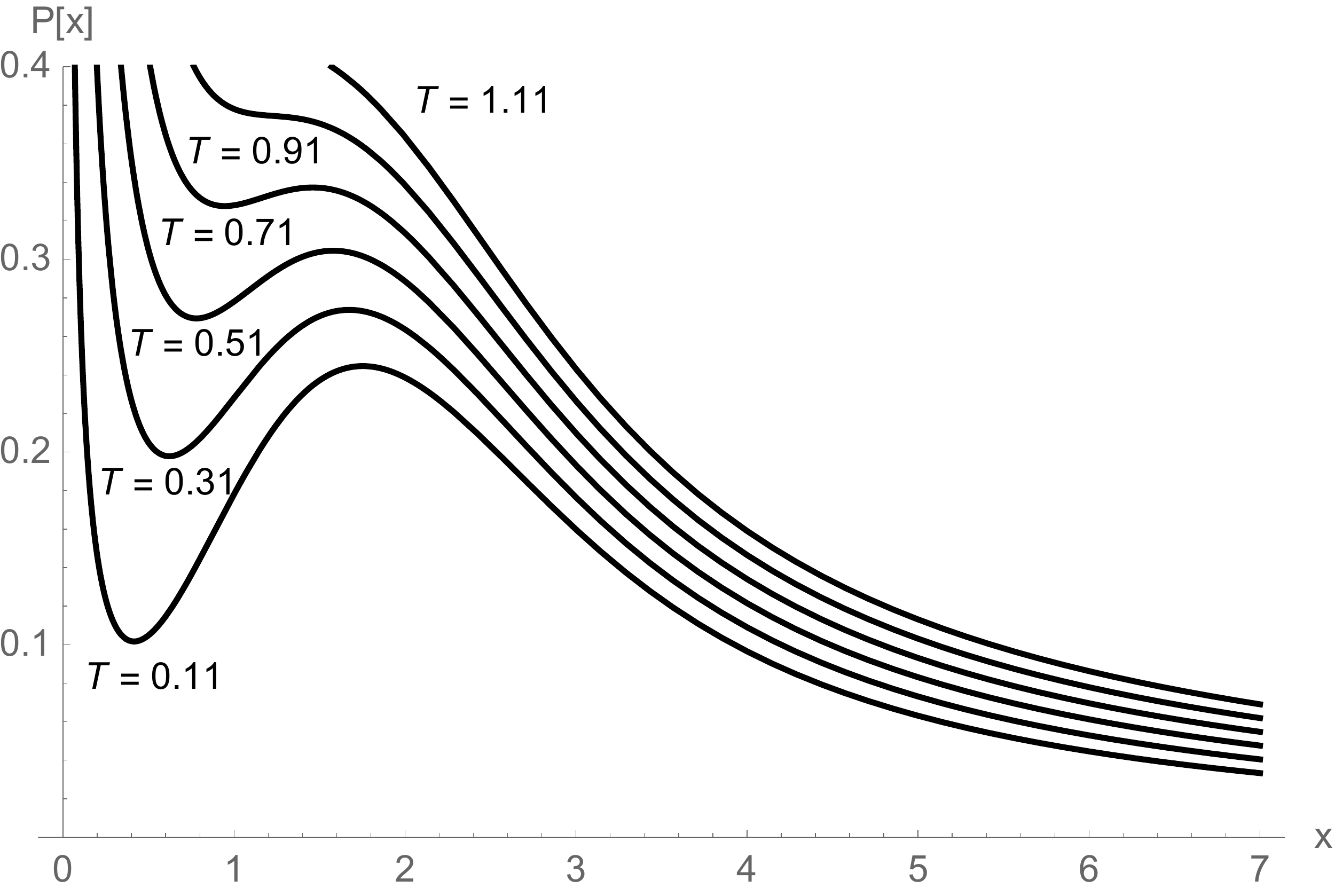} 
\end{center}
\caption{\small Pressure from \eqref{reg_press}. Here $v = 4Gr_ +$, $x = r_ +/a$, $a^2  = 0.8$, $Gq^2=2.88$ and $G=r_0=1$. 
\label{FIGb}}
\end{figure}

As in the previous section we can study the fluid dual to the black hole. By introducing the vacuum pressure $P=-\rho_{\rm AdS}$, we obtain the following equation of state
\begin{equation}
P=\frac{T_{\rm H}}{v}+\frac{2G^2 q^2}{\pi^2}\frac{F'(v)}{v}
\end{equation}
We notice that the function $F(v)$ can provide a different phase structure. By using \eqref{eq:regmetric2}, we can write
\begin{equation}
P=\frac{T_{\rm H}}{v}+\frac{2G^2 q^2}{\pi^2}\ \frac{\gamma^2\left(1; \frac{v^2}{64G^2l^2}\right)}{v^2}
\label{reg_press}
\end{equation}
whose plot is given in Fig. \ref{FIGb}. The equation of state turns out to be of the van der Waals type. At high temperatures, \textit{i.e.}, large horizon radii, the system approaches a perfect gas behavior.  By lowering the temperature, however, the typical S type profile indicated a mixture of phases during the transition from a low compressibility configuration (small black hole) to a progressively higher compressibility configuration (bigger black hole). 
Such a behavior is in agreement with the findings about phase transitions of regular black holes in $(3+1)$-dimensions \cite{NiT11,Tzikas2019}.

\section{Conclusions}
\label{sec:concl}

In this paper we faced the problem of the three dimensional charged static black hole solution. After presenting the standard BTZ geometry we highlighted three main features: the presence of a conical singularity, the arbitrariness of the mass parameter and the difficulty in defining the  internal energy. We also showed that such characteristics are connected to the presence of a logarithm as a solution of the $2+1$ dimensional Poisson equation.

Given this background we proposed to exploit the arbitrariness of the length scale entering the argument of the logarithm as well as the integration constant of the Einstein equations to extend the parameter space conventional BTZ solution. We showed that, despite the mass term has disappeared from the metric coefficient, it is still possible to define an internal energy consistent with the Area Law. Our method differs from what already presented in the literature about the introduction of a large box radius to control the logarithmic term \cite{MTZ00,CaM09}. We showed that such methods  describe the physics equivalently. As a piece of evidence, we established a formal duality  black-hole/fluid of the same kind of what found in the literature for black holes in AdS vacuum \cite{GKM12}. Interestingly, the BTZ solution behaves like a perfect gas with a short scale correction due to the electrostatic interaction.


In the second part of the paper, we set the parameters of the solution to eliminate the monopole term and the related conical singularity.  We showed that such a procedure can occur owing to the extension of the parameter space previously derived. The BTZ geometry on the other hand cannot reproduce the Minkowski limit since it permanently affected by a conical singularity. We also show that the necessity of a negative cosmological constant for horizon formation is a characteristic of the BTZ parametrization. In general $2+1$-dimensional black hole exists also in a de Sitter background space.

 As a further improvement,  we introduced a ultraviolet cutoff to amend the short scale behavior of the electrostatic potential.
 In such a way, we derived  a short scale singularity free solution, that admits horizons and does not require any modification of the action, as previously claimed \cite{Cle95}. The related thermodynamics discloses a new phase structure that resembles that of a real gas. Physically this means that the short scale cut off behaves like the parameter controlling the excluded volume in a Van der Waals gas.

\subsection*{Acknowledgements}
One of us (P. G.) was partially supported by Fondecyt (Chile) grant 1180178 and by ANID PIA / APOYO AFB180002. P.G. also wishes to thank the Abdus Salam ICTP for hospitality.
The work of P.N. has been partially supported by the grant NI 1282/3-1 of the project 
``Evaporation of microscopic black holes" of the German Research Foundation (DFG), 
by the Helmholtz International Center for FAIR within the framework of the LOEWE 
program (Landesoffensive zur Entwicklung Wissenschaftlich-\"{O}konomischer Exzellenz) launched by the State of Hesse and by GNFM, the Italian National Group for Mathematical Physics.
The authors thank A. G. Tzikas for valuable comments on the manuscript.

\end{document}